\documentclass[final,3p,times,twocolumn,sort&compress]{elsarticle}
\usepackage{epic,eepic}

\usepackage{graphicx}

\usepackage{amssymb}
\usepackage{amsmath}

\usepackage{xspace}

\usepackage{courier}
\usepackage{helvet}
\usepackage[slantedGreek]{mathptmx}

\usepackage[dvips]{hyperref}

\newcommand{\labr}{LaBr$_3$:Ce\xspace}

\hypersetup{pdftitle={High count rate gamma-ray spectroscopy with Lanthanum Bromide scintillation detectors},
            pdfauthor={Bastian Loeher},
}

\begin{document}

\begin{frontmatter}



\title{High count rate $\gamma$-ray spectroscopy with \labr scintillation detectors}


\author[label1,label2]{B.\ L\"oher\corref{cor}}
\ead{b.loeher@gsi.de}
\author[label1,label2,label3]{D.\ Savran}
\author[label1,label2]{E.\ Fiori}
\author[label4]{M.\ Miklavec}
\author[label5]{N.\ Pietralla}
\author[label4]{M.\ Vencelj}
\address[label1]{ExtreMe Matter Institute EMMI and Research Division, GSI Helmholtzzentrum f\"ur Schwerionenforschung, Planckstr. 1, 64291 Darmstadt, Germany}
\address[label2]{Frankfurt Institute for Advanced Studies FIAS, Ruth-Moufang-Str. 1, 60438 Frankfurt am Main, Germany}
\address[label3]{Innovation Centre for Advanced Sensors and Sensor Systems, INCAS$^{3}$, Assen, The Netherlands}
\address[label4]{Jo$\check z$ef Stefan Institute, Jamova cesta 39, 1000 Ljubljana, Slovenia}
\address[label5]{Institut f\"ur Kernphysik, TU Darmstadt, Schlossgartenstr.\ 9, 64289 Darmstadt, Germany}

\cortext[cor]{Corresponding address: ExtreMe Matter Institute EMMI and Research Division, GSI Helmholtzzentrum f\"ur Schwerionenforschung, Planckstr. 1, 64291 Darmstadt, Germany, Tel.: +49 6159711803; fax: +49 6159713475}

\begin{abstract}

The applicability of \labr detectors for high count rate $\gamma$-ray spectroscopy is investigated.
A 3"x3" \labr detector is used in a test setup with radioactive sources to study the dependence of energy resolution and photo peak efficiency on the overall count rate in the detector.
Digitized traces were recorded using a 500~MHz FADC and analysed with digital signal processing methods.
Good performance is obtained using standard techniques up to about 500~kHz counting rate. A pile-up correction method is applied to the data in order to further improve the capabilities at even higher rates with a focus on recovering the losses in efficiency due to signal pile-up.
It is shown, that $\gamma$-ray spectroscopy can be performed with only moderate lossen in efficiency and high resolution at count rates even above 1~MHz and that the performance can be enhanced in the region between 500~kHz and 10~MHz by using the applied pile-up correction techniques.

\end{abstract}

\end{frontmatter}

\section{Introduction}\label{sec:intro}

In modern nuclear physics experiments, $\gamma$-ray spectroscopy is one
of the major tools to investigate the structure of stable and
unstable nuclei. The detection and spectroscopy of
the emitted photons in the de-excitation of particle-bound excited
states is a mandatory ingredient in many studies of nuclear reactions.
Consequently $\gamma$-ray detectors
play an important role in the configuration of experimental setups.

Different types of detectors are available.
The optimum choice depends on the requirements of the $\gamma$-ray
spectroscopy experiments.
High-Purity Germanium (HPGe) detectors are often used for highest
resolution, while several
scintillator based detectors provide better timing properties
and efficiencies at lower costs and reduced resolution.

In many experiments, the emitted $\gamma$ rays in the reaction
of interest are accompanied by a rather large $\gamma$-ray background.
This radiation originated from photons produced in background
reactions such as bremsstrahlung by fast secondary electrons or
reactions within the beam dump or other material along the beamline.
To avoid signal distortion, the maximum count rate capability of each detector must not be exceeded. As a consequence the rate in the photon detectors
often limits the beam intensity and therefore the reaction rate.

The recently developed scintillator material lanthanum bromide (\labr, tradename BrilLanCe\textsuperscript{\texttrademark}380 by Saint Gobain \protect{\cite{SGo08}}) is well suited to push the limitations posed by this problem.
Its excellent timing properties (down to 100~ps in optimal conditions \cite{ilti2006,mosz2006b,zhu2011}), temperature stability \cite{loef2002,mosz2006} and high energy resolution for a scintillator (3\% at 662~keV \cite{quar2007,paus2007}), which is dominated by statistical contributions \cite{deva2006}, have made \labr the material of choice for many nuclear physics experiments including $\gamma$-ray spectroscopy 
\cite{smit2008,regi2010},
medical imaging 
\cite{mose2005,kuro2010}
or industrial applications \cite{kuli2007,milb2007}.
The high light yield (165\% of that of NaI) and the very short scintillation light decay constant (between 20 and 30~ns \cite{loef2002,mosz2006}) pose \labr as a well suited candidate material to allow $\gamma$-ray spectroscopy at very high count rates.
This property has been briefly discussed by comparing \labr to NaI \cite{ilti2006}.
However, the investigation in \cite{ilti2006} does not include the study of efficiency and resolution -- the key properties for $\gamma$-ray spectroscopy in nuclear physics experiments -- when increasing the count rate in the \labr detector.

In this article, we investigate in detail the high count rate capabilities of \labr and particularly the above mentioned energy resolution and photo peak efficiency with respect to the count rate.
Further improvement is achieved by applying digital signal-processing (DSP) techniques in order to correct signal distortions and restore efficiency losses due to pile-up.
It is shown that high-resolution $\gamma$-ray spectroscopy with \labr detectors is possible at count rates well in excess of 1~MHz.
Simulations with synthetic signals have been carried out to investigate the limits of this approach, and to support the experimental results.

Section~\ref{sec:setup} of this article will shortly introduce the experimental setup that was used for the measurements,
while the data analysis, including feature extraction from the digitized signals, are detailed in Section~\ref{sec:ana}.
This section also describes how a timestamp-based pile-up correction method, first introduced in \cite{Ven09}, can be used to further improve the performance of \labr detectors at high count rates.
In Section~\ref{sec:results} the results of these investigations will be shown and an outlook is presented in Section~\ref{sec:concl}.


\begin{figure}[!t]
\centering
\includegraphics[width=0.4\textwidth]{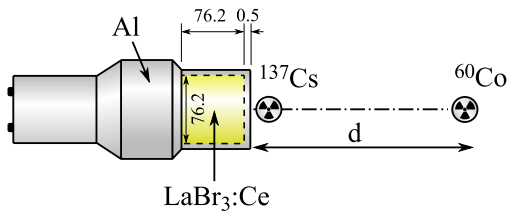}
\caption[Experimental setup]{
A schematic illustration of the experimental setup (dimensions in mm).
The weaker $^{137}$Cs source was kept at a fixed distance to the detector (close to the aluminum front cap to get the highest count rate possible),
while the distance of the strong $^{60}$Co was varied to create a variable amount of background radiation.
}\label{fig_setup}
\end{figure}

\section{Detector and Setup}\label{sec:setup}

The experimental setup for the present investigation is shown in Fig.~\ref{fig_setup}.
The large volume 3"x3" \labr detector was irradiated by two different radioactive sources.
One $^{137}$Cs source of medium activity (45~kBq) was placed in contact with the front face of the detector,
while a $^{60}$Co source with high activity (73~MBq) was placed at a variable distance $d$.
The $^{137}$Cs source was fixed in place for all measurements concerning the efficiency studies and thus provides the stable reference,
while the strong $^{60}$Co source was moved within a range of 5-100~cm in order to vary the overall count rate.
The count rate using only the $^{137}$Cs source was 10~kHz.
Bias voltage for the Hamamatsu R10233 photomultiplier tube (PMT) was delivered from an \textit{iseg} NHQ-203M power supply through a model 2007 resistive voltage divider base from Canberra and operated at bias voltages from 300~V to a maximum of 1200~V.
The anode signal of the PMT was amplified by a fast timing filter amplifier (TFA) with 1~$\mu$s differentiation and 1~ns integration time settings in oder to match the dynamic range of the sampling ADC.
The amplified signal was then recorded with a Struck SIS3350 digital sampling ADC with a sampling rate of 500~MHz, a bandwidth of 250~MHz and a resolution of 12~bit \cite{SIS35}.
The length of each signal trace was set at $10^5$ data points corresponding to $0.2$~ms of continuous samples.
To reach a live time of $2$~s, in total $10^4$ traces have been recorded for each distance $d$.
Data acquisition was done using a Struck SIS3104/SIS1100e PCIe-VME-interface and a standard PC running the GECKO data acquisition software\cite{gecko} with online signal processing.
Using the PCIe-VME-interface high data transfer rates of more than $100$~MB/s between the VME electronics and the PC could be reached to allow fast readout of the sampling ADC.

\section{Signal analysis and pile-up correction}\label{sec:ana}
The information conveyed in the digitized signals has to be extracted using DSP methods that are described in this section.
At first the timestamp is defined from the fast rising edge of the pulse,
because the steepest part of the slope conveys precise timing information.
Afterwards regions of the trace without any signals are identified based on the timestamps.
Then the baseline offset is corrected.
The energy deposit information can then be extracted from the integral over each signal.
These parts of the data analysis are described in detail in the following.

\begin{figure}[t]
\centering
\includegraphics[width=0.5\textwidth]{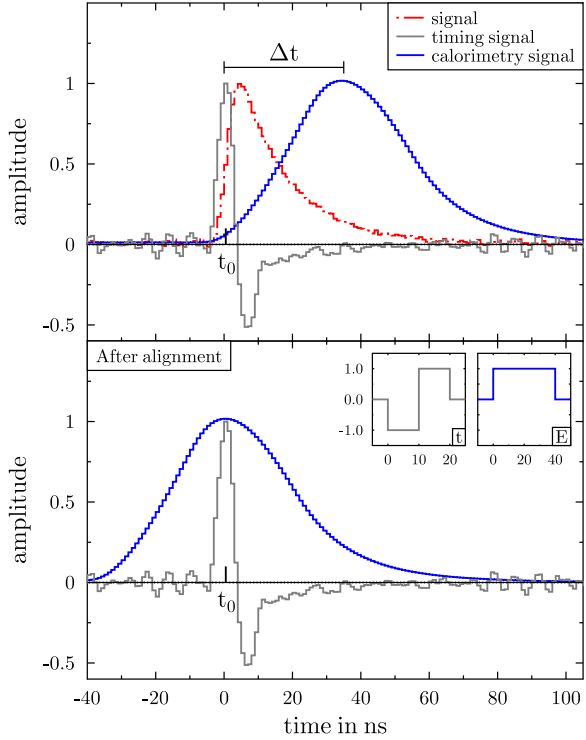}
\caption[Digital calorimetry]{
\textit{Top:} Illustration of the applied signal analysis for a single pulse with a rise time of 10~ns and a decay constant of 25~ns.
A differentiating filter (t) transforms the signal (dashed, red line) into the timing signal (grey).
After filtering with an integrating box filter (E), the signal is transformed into a calorimetry signal (solid, blue).
The delay $\Delta t$ between these two signals is independent of the amplitude.
The trigger time $t_0$ is determined from the timing signal.
\textit{Bottom:} After taking into account the delay $\Delta t$, the amplitude of the signal is determined at the time $t_0$.}\label{fig_filter}
\end{figure}

\subsection{Feature extraction}\label{subsec:feat}
Several different steps of signal analysis are performed on the data in order to extract the basic signal information (i.e. energy and timing).
The baseline-offset voltage is estimated by averaging over 10 data points of the signal trace in regions where no $\gamma$-ray signal is present and is then subtracted from the original signal.
These regions are identified by requiring a minimum distance of 120 ns after a trigger and a minimum distance of 60 ns before a trigger, with a trigger threshold of 3$\sigma$ of the noise level.
At very high count rates the probability for such "empty" regions in a trace decreases, such that the determination of the baseline becomes more difficult.
At the cost of precision it is still possible to derive a value for the baseline, if the trace contains at least one of such regions.
If no suitable regions are present, then different methods have to be employed,
which however was not necessary in the present case.
The dashed red line in Fig.~\ref{fig_filter} shows the data for one typical signal trace with the baseline offset voltage already subtracted.
To derive a reference point in time (trigger time or timestamp) from this signal,
the data are convoluted with a discrete bipolar Finite Impulse Response (FIR) filter kernel with a width that equals the signal rise time of 10~ns, which is independent of the amplitude, because the signal shape does not depend on the energy.
This filter kernel has the value $-1$ for the first 10~ns and the value $+1$ for another period of 10~ns (see inset "t" in the lower panel of Fig.~\ref{fig_filter}).
The resulting bipolar timing signal is shown in gray.
From the maximum, defined by the data point with the highest amplitude, of this timing signal it is possible to derive an accurate trigger timestamp,
which is denoted as $t_0$ in the illustration.

In the next step the deposited energy is determined by integrating the signal.
The original signal is filtered with a unipolar FIR filter kernel, i.e. its value is $+1$ over its duration, with a width that ideally equals the length of the signal (shown with a width of 40~ns in inset "E"),
in order to produce an integrated calorimetry signal (solid blue).
The amplitude of this signal corresponds to the amount of deposited energy.
For computational reasons the width of the filter can be reduced at the cost of precision.
In the present analysis a width of 40~ns was used for all of the measurements.
The resulting timing and calorimetric signals are aligned in time by shifting them with respect to each other, as denoted in Fig.~\ref{fig_filter} by $\Delta t$.
The value of this delay is constant for any recorded signal,
as long as the signal shape is constant, which is the case in the present study.
The energy deposit is thus given by the value of the calorimetry signal at $t = t_0 + \Delta_t$.
This approach for extracting the signal properties has been chosen for two reasons:
First this is an exceedingly simple and numerically stable approach,
and more importantly this approach allows for a correction of the measured amplitudes in the case of strong pile-up, as described in the following.

\begin{figure}[t]
\centering
\includegraphics[width=0.5\textwidth]{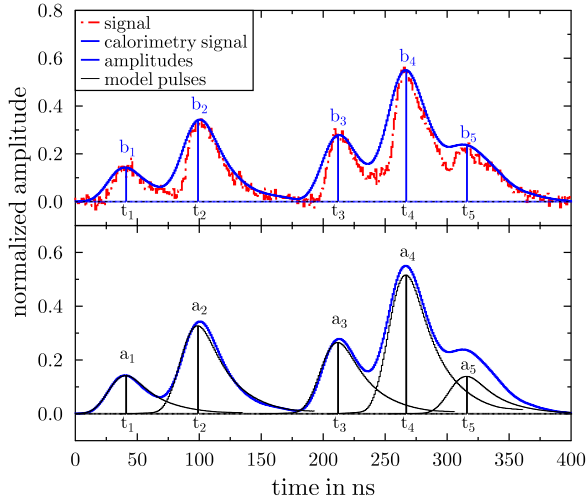}
\caption[Pile-Up Correction]{\textit{Top:} Example of a pile-up situation.
Amplitudes (blue) have been measured from the calorimetry filtered signal (solid line).
\textit{Bottom:} After pile-up correction the measured amplitudes $b_i$ are corrected to yield the true amplitudes $a_i$ (black).
The superposition of the model pulse shapes (thin lines) results in the filtered signal (thick line).
}\label{fig_puc}
\end{figure}

\subsection{Pile-Up correction}\label{subsec:puc}
At increasing count rates the average time difference between two consecutive pulses will be shorter than the individual pulse length. Consequently different pulses overlap in time, resulting in a correlation of their properties, which -- if not corrected for -- finally leads to a reduction of the energy resolution and especially to a loss of photopeak efficiency.
The pile-up correction method that was used in the analysis has first been presented in~\cite{Ven09}.
The detailed description can be found there, and only a short outline is given in this article.

This method relies on the assumption that the pulse shape of every signal from the same scintillation detector is stable and independent from its amplitude,
so that the complete signal trace can be expressed as a linear combination of single pulses, as illustrated in Fig.~\ref{fig_puc}.
This assumption is sufficiently well fulfilled for \labr detectors.
The complete trace can thus be described by the timestamp of each pulse and the corresponding pulse amplitude,
as well as a single model pulse shape common to all pulses that characterizes the dynamics of a single signal.
Based on this model the correlation of neighboring signals can be resolved in order to restore the correct $\gamma$-ray energy information from the measured amplitudes.

Figure~\ref{fig_puc} illustrates the principle of the applied pile-up correction method.
A signal trace (dashed red) with five signals from the \labr detector is shown at the top together with the computed calorimetry signal (blue solid).
For illustration purposes the time delay introduced by the digital filters is removed as described above.
The single signals partly overlap each other,
such that the amplitudes $b_i$ taken from the calorimetry signal at times $t_i$ signal are overestimated
(especially visible in the case of $b_4$ and $b_5$).
The solution to correct for this overestimation is to treat each of the amplitudes $b_i$ as a superposition of the true amplitude $a_i$ of the corresponding signal and additional contributions $c_i$ from the surrounding signals:

\begin{equation}
b_i = a_i + \sum^{N}_{\substack{j = 0  \\ j\ne i}}c_j.
\end{equation}
The contribution of the $j$th neighbor signal can be expressed as the true amplitude $a_j$ of the neighbor signal weighted with a factor $m_{ij}$,
such that $c_j = m_{ij} \, a_j$.
The value of this factor depends solely on the time difference $t_{ij} = t_i - t_j$ between the time of signal $i$ and the time of the neighbor $j$ and can be derived with the use of the model pulse shape $m(t)$.
Taking the value of the model pulse shape at the time $t_{ij}$ with respect to the trigger time $t_0$ of the model pulse yields the desired weight $m_{ij} = m(t_0 + t_{ij})$, resulting in

\begin{equation}
b_i = \sum^{N}_{j = 0} m_{ij} \, a_j = \sum^{N}_{j=0} m(t_0 + t_{ij}) \, a_j.
\end{equation}
For all signals in the trace this equation can be written as a set of linear equations

\begin{equation}
\mathbf{b} = \mathbf{M\,a},
\end{equation}
that can be solved by inverting the matrix $\mathbf{M}$ containing the weights $m_{ij}$ and thus produce a solution for the true amplitudes $a_i$

\begin{equation}
\mathbf{a} = \mathbf{M^{-1}\,b}.\label{equ_inverse}
\end{equation}
The model pulse shape is determined using data recorded with a low count rate, in order to avoid pile-up situations.
From this data a large number of single signals 
are extracted.
These signals are averaged and normalized to amplitude 1.0 to produce the model pulse shape.

The bottom part of Fig.~\ref{fig_puc} shows the same calorimetry signal again, but now together with the computed amplitudes $a_i$ and the corresponding constituent signals (black).
These constituent signals are produced by scaling the model pulse shape with the true amplitude $a_i$ as determined using Eq.~\ref{equ_inverse} and shifting it in time to match the trigger time $t_i$.
From this figure it is obvious, that the calorimetry signal results from a superposition of all constituent signals.
The free parameters of this algorithm are the extents of the matrix $\mathbf{M}$,
which control how many neighbor signals to take into account for each calculation.
It is reasonable to assume a small number of rows and columns,
because the model pulse shape only has non-zero elements within about 200~ns around its trigger time.
This fact allows to calculate the matrix inversion with only a small amount of computing resources,
and even enables the implementation of the algorithm in hardware for real-time applications (e.g. FPGA or DSP).
In the present study the matrix inversion has been carried out for each event individually. This can be done efficiently in hardware as well. For the case of only one neighboring signal on each side the inversion is simple, but for higher order corrections a more specialized method as described in \cite{nova2009} is necessary.

\subsection{Application limits}\label{subsec_app_lim}

\begin{figure}[t]
\centering
\includegraphics[width=0.5\textwidth]{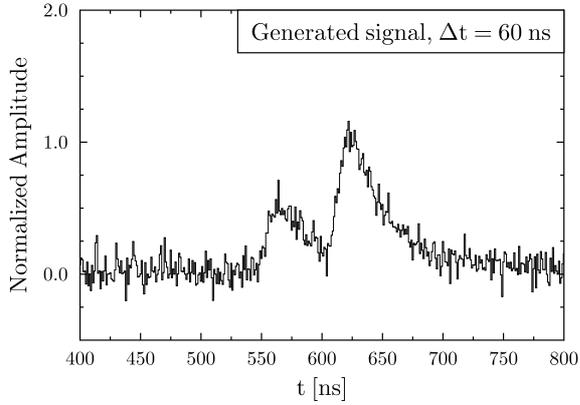}
\caption[Limit of Pile-Up Correction]{
Generated signal group with a distance of $\Delta t = 60$~ns between the two signals with an amplitude ratio of 1:2. 
}\label{fig_limit_of_puc2}
\end{figure}

\begin{figure}[t]
\centering
\includegraphics[width=0.5\textwidth]{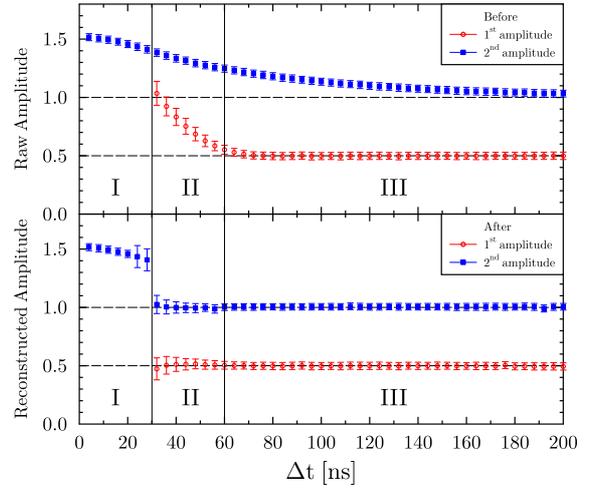}
\caption[Limit of Pile-Up Correction Result]{
\textit{Top:}
Measured raw amplitudes ($b_i$) with respect to time difference $\Delta t$ (first signal with squares,
second with circles).
Horizontal lines indicate "true" generating amplitudes.
Uncertainties in each point result from statistical errors.
Three distinct ranges in $\Delta t$ can be identified: 
(I) Signals are indistinguishable
(II) Incorrect amplitudes due to pile-up
(III) First amplitude is measured correctly, but second one is still overestimated.
\textit{Bottom:} 
Reconstructed amplitudes ($a_i$). 
Three distinct regions can be identified:
(I) Trigger limited mis-identification, 
(II) Transition region with increased uncertainty on the result, 
(III) Incorrectly measured amplitudes are accurately restored.
}\label{fig_limit_of_puc}
\end{figure}

To study the performance of the digital signal analysis and the limits for the application of the pile-up correction algorithm,
test cases have been constructed with two synthetic signals at a well defined distance $\Delta t$.
Fig.~\ref{fig_limit_of_puc2} shows an example with a distance of $\Delta t = 60$~ns and an amplitude ratio of 1:2 of the two signals (the second is twice as large as the first).
The signals are generated from an exponential function with a decay constant of 60~ns, which is smoothed with a Gaussian filter to create a finite rise time of about 10~ns (matching the properties of real detector signals) and then scaled to its amplitude.
White Gaussian noise is added to the trace to produce a signal-to-noise ratio (SNR) of 10 for the first and 20 for the second signal, respectively (SNR is here defined as pulse height divided by the root mean square of the noise level).
This amount of noise is chosen to simulate small detector signals.
Based on a typical noise level of 10~mV$_{rms}$ (root mean square), the simulated pulse heights correspond to 100 and 200 mV, respectively.
These pulse heights correspond to an energy deposition on the order of 100 and 200~keV for the experimental settings used, and thus represent fairly small signals.
For these small signals the limitations of the method with respect to timing are more visible, compared to large signals.
The parameter $\Delta t$ is varied from 0 to 200~ns in steps of 4~ns,
and for each step a set of 5000 traces
was generated.
The traces were analysed with the approach described in Sec.~\ref{subsec:feat} to determine the trigger times $t_i$ and the raw amplitudes $b_i$ of the two signals.
For each $\Delta t$ a gaussian distribution was fitted to the resulting histograms to determine the mean value and the FWHM of each of the two peaks.

The results are shown in the upper part of Fig.~\ref{fig_limit_of_puc} as a function of the time difference $\Delta t$ of the two generated signals.
The error bars for each point are drawn according to the determined FWHM.
Three distinct ranges in $\Delta t$ can be identified: 
(I) The time difference is comparable to or even smaller than the rise time of the signals. They appear as one single signal and the used trigger algorithm cannot distinguish between them; 
(II) At intermediate time distances, both amplitudes are incorrectly measured, because of pile-up. This region extends to about three times the rise time for our simulation; 
(III) The first amplitude is measured correctly, because in this region the influence of the rising edge of the second signal on the amplitude of the first one becomes negligible. The amplitude of the second signal is still overestimated (top of Fig.~\ref{fig_limit_of_puc}).

The lower part of Fig.~\ref{fig_limit_of_puc} shows the results once pile-up correction method has been applied.
For this simulation the method only takes into account the next neighbors of each signal (1$^{st}$ order), because only two signals participate.
The three different regions can be identified here as well:
Region (I) is the region of unresolvable pile-up,
where the pile-up correction cannot be applied,
because the two signals cannot be separated based on the current analysis.
For the same reason it is not possible to reject these signals,
based purely on the trigger information.
Region (II) is the transition region, where the pile-up correction method can be applied, and yields the proper result.
However, for shorter time intervals the statistical uncertainty increases.
The magnitude of this increase is on the order of 2.5-3.0 times the intrinsic uncertainty.
Depending on the application, the events in this transition region can be identified and rejected.
Besides depending on the detector resolution, these signals may be corrected well enough to contribute to the photo peak.
The amplitudes in region (III) can be properly restored within their intrinsic uncertainties.

The simulation shows that in a range of 60-200~ns for $\Delta t$, the simulated pile-up correction method yields a good reconstruction of the signal amplitude,
while the uncorrected amplitude of the second signal is determined incorrectly. 
When applying the same method to real data an improvement of the extracted properties is expected for count rates where a large fraction of the signals have a time difference in this range.
At lower or higher count rates the improvement is expected to be much smaller, because either an insignificant amount of signals overlap or because the average distance between two signals is too small to allow for precise determination of the timestamp.

\section{Experimental results}\label{sec:results}
In this section the experimental results from the measurements detailed in Sec.~\ref{sec:setup} are presented.

\begin{figure}[t]
\centering
\includegraphics[width=0.5\textwidth]{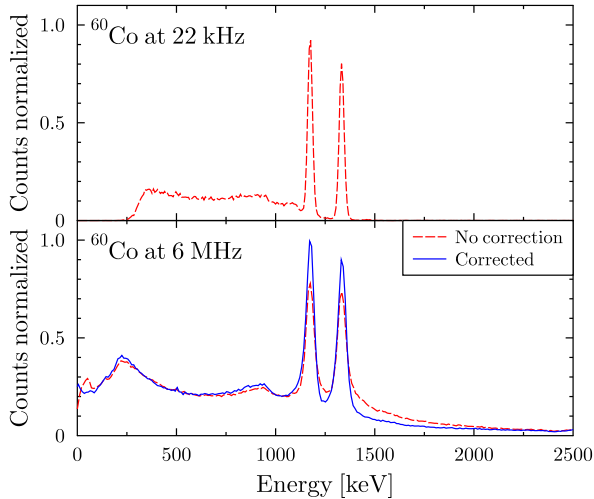}
\caption[$^{60}$Co spectra]{
(Color online) Energy spectra of emitted $\gamma$-rays from a $^{60}$Co source recorded with the \labr detector.
\textit{Top:}
Large distance (1.5~m) between source and detector to induce a low count rate of 22~kHz.
\textit{Bottom:}
The count rate increases to 6~MHz at smaller distances (20~cm).
Pile-up events are visible above energies of 1.3 MeV in the spectrum without correction (dashed red).
These are partly restored in the corrected spectrum (solid blue).
}\label{fig_60co_puc}
\end{figure}

\begin{figure}[t]
\centering
\includegraphics[width=0.5\textwidth]{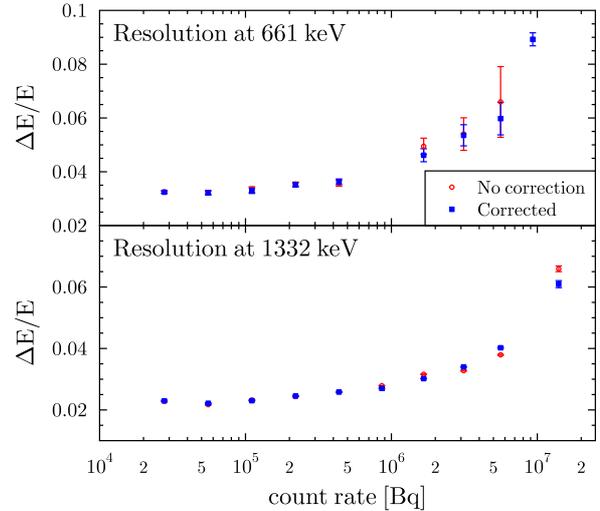}
\caption[Energy Resolution]{
\textit{Top:}
Relative resolution at 661~keV ($^{137}$Cs) without correction (red circles) and with correction (blue squares). At the highest count rate the resolution without correction could not be determined.
\textit{Bottom:}
Relative resolution at 1332~keV ($^{60}$Co).
}\label{fig_resolution}
\end{figure}

\begin{figure}[t]
\centering
\includegraphics[width=0.5\textwidth]{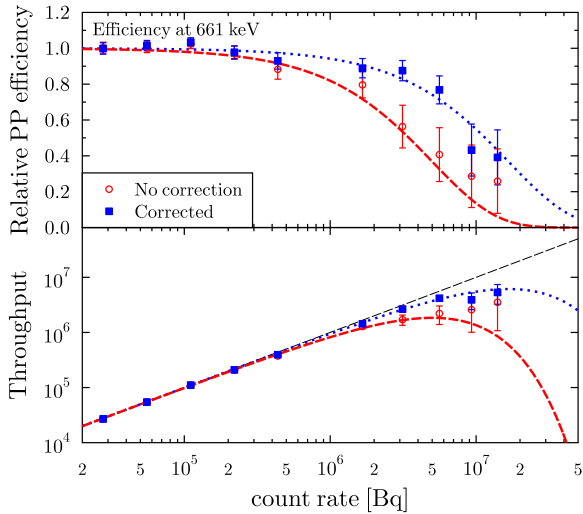}
\caption[Efficiency and Throughput]{
\textit{Top:}
Relative photo peak efficiency at 661~keV normalized to the value at the lowest count rate (red circles without correction, blue squares with correction).
The curves indicate the probability for detecting only one signal within a time interval $\tau$ (dotted: $\tau = 60$~ns, dashed: $\tau = 200$~ns) at the respective count rate.
\textit{Bottom:}
Throughput (product of relative PP efficiency and count rate).
At count rates above 1~MHz a clear enhancement is visible after correction.
The curves are calculated from the above probability characteristic.
}\label{fig_efficiency}
\end{figure}

Figure~\ref{fig_60co_puc} shows energy spectra of emitted $\gamma$-rays from the strong $^{60}$Co source recorded with the \labr detector.
The top part shows the spectrum taken with a low count rate, while the source was placed at a distance of 1.5~m to the detector.
The lower part of Fig.~\ref{fig_60co_puc} shows the spectrum when recorded with the source at a distance of 20~cm.
This new situation translates in a considerable increase in count rate from 20~kHz to 6~MHz and makes the pile-up effects apparent in this spectrum, which can be seen as high energy tails of the peaks and the increase of events with energies much larger than 1.33~MeV.
Without pile-up correction the red (dashed) spectrum has less photo peak efficiency (5\% peak-to-total compared to 16\% at 20~kHz) and also the energy resolution is worse when compared to the spectrum at low count rates.
After pile-up correction taking into account up to three neighboring signals on each side (blue) the photo peak efficiency is partly recovered.
At the same time the number of pile-up events in the energy region from 1.4 MeV to 2.5 MeV is reduced by 38\%, compared to the spectrum without pile-up correction.

In the following the recovery of the efficiency using the pile-up correction is analysed quantitatively using in addition to the $^{60}$Co source the $^{137}$Cs source, which provides a constant count rate (reference).
The photo peak efficiency is determined from the peak area of the 661~keV peak corresponding to the energy of the $\gamma$-rays emitted by the $^{137}$Cs source.
The peak region has been fixed for all count rate settings.
The obtained values are then normalized to the peak area at the lowest count rate of 28~kHz.

The results on the dependence of the energy resolution and photo peak efficiency on the count rate are presented in Figs.~\ref{fig_resolution} and \ref{fig_efficiency}, respectively.
The energy resolution stays below 4\% at 661~keV for count rates below 500~kHz, and does not increase above 10\% for the highest count rate measured (14~MHz). The same is true for the resolution at 1332~keV. From Fig.~\ref{fig_resolution} it is also clear that the pile-up correction (blue squares) does not restore the energy resolution by a significant amount.
The lengths of the error bars correspond to statistical errors only.
For the highest count rate the uncorrected spectrum could not be analysed.

In the upper part of Fig.~\ref{fig_efficiency} the results for the relative photo peak efficiency at 661~keV is shown.
Again red circles indicate the efficiency before pile-up correction, and blue squares mark the results after pile-up correction.
Along with the data points two different curves are drawn.
These curves correspond to the probability that within a certain time interval $\tau$ a second following signal occurs, if a random sequence of signals is assumed.
These probabilities are calculated using a Poisson distribution, according to

\begin{equation}
P(N,n)=\frac{(n \tau)^N e^{-n \tau}}{N !},
\end{equation}
where N denotes the number of signals in the time interval of length $\tau$ and $n$ the mean number of pulses per time interval (count rate).
Thus, $P(0,n)$ is the probability for a second pulse occuring within the time $\tau$ for a given count rate $n$.
For the illustration two different values for $\tau$ (blue: $\tau = 60$~ns, red: $\tau = 200$~ns) as estimated from the simulation in Sec.~\ref{subsec_app_lim} are used.
The measured data follow the shape of the two curves in both cases, which indicates that the simulation correctly predicts the region where the application of the pile-up correction method will improve the results.
In the range between 500~kHz and 10~MHz the advantage from using the pile-up correction method is most noticeable.
For lower count rates the effect of pile-up can be neglected, while for higher count rates the leading edges of the pulses overlap and thus prohibit accurate triggering.
The lower part of Fig.~\ref{fig_efficiency} shows the resulting estimated throughput (given by the product of relative efficiency and the count rate) together with the ideal case where the throughput equals the incoming count rate (black line) and the two curves from the theoretical prediction.
This figure demonstrates, that at a low count rate the effect of pile-up correction is negligible, while at count rates above 1~MHz the improvement is significant, increasing the ideal count rate and the maximum throughput by a factor of 3.


\section{Conclusions and Outlook}\label{sec:concl}
The analysis has shown that \labr detectors are suitable for $\gamma$-ray spectroscopy even at very high count rates, exceeding 1~MHz.
Using the presented pile-up correction method, the loss of photo peak efficiency from piled-up pulses can be alleviated.
A reasonable range for the application of this method is between 500~kHz and 10~MHz, corresponding to an average distance of two signals in the range from 60 to 200~ns.
At count rates above 10~MHz the analysis of the digitized signals becomes more demanding, because additional effects decrease the accuracy of feature extraction.
One of these effects is the saturation of the PMT with increasing count rate,
which can be avoided either with a lower bias voltage or by using an active voltage divider with transistor stages, which, however was not available for the present study.
The fact, that the applied pile-up correction method has only little influence on the resolution points to the PMT and the voltage divider as the dominant factor for the energy resolution in our setup.
The finite rise time of the used PMT in combination with the 500~MHz sampling rate of the ADC limits the precision of the trigger time determination and thus also the accuracy of the reconstructed amplitude.
This is most notable, when the time difference between two pulses is small and a small change in time results in a large change in amplitude.
Other effects that arise at these high count rates like unresolvable pile-up and the fact that the baseline offset can not be determined in a simple manner need additional investigation.


\section*{Acknowledgments} 
The authors thank H.~W\"ortche (INCAS$^3$), and M.~Kirsch (Struck Innovative Systeme) for valuable and inspiring discussions.
B.L. and D.S. thank the Institut Jo\v{z}ef Stefan, Ljubljana for their kind hospitality.
This work was supported
by the Alliance Programme of the Helmholtz Association (HA216/EMMI),
by the LOEWE programme of the state of Hesse through the Helmholtz International Center for FAIR, 
by the Deutsche Forschungsgemeinschaft (SFB 634)
and the Innovation Centre for Advanced Sensors and Sensor Systems, INCAS$^{3}$, Assen, The Netherlands.
Further support was provided
by the Slovene Competence Center for Biomedical Engineering,
and by Instrumentation Technologies of Solkan, Slovenia.


\bibliographystyle{apsrev}
\bibliography{ikpbib_nourl}

\end{document}